\documentclass[12pt]{article}

\include{amssym}

\def\mathbf{\vec}

\def\ii{\'{\i}}

\newcommand{\njl}{\mathrm{NJL}}
\newcommand{\crit}{\mathrm{crit}}
\newcommand{\e}{\mathrm{E}}
\newcommand{\st}{\mathrm{st}}
\newcommand{\mix}{\mathrm{mix}}
\newcommand{\ud}{\mathrm{d}}
\newcommand{\p}{\mathrm{p}}

\begin{document}

\centerline{\Large\bf  
           The 't Hooft determinant resolution}
\vspace{0.2cm}
\centerline{\Large\bf  of the $\eta'$ puzzle}
\vspace{1cm}

\centerline{\large Alexander A. Osipov\footnote{Joint Institute for 
            Nuclear Research, Laboratory of Nuclear Problems, 
            141980 Dubna, Moscow Region, Russia. 
            Email address: osipov@nusun.jinr.ru}, 
            Alex H. Blin\footnote{Email address:
            alex@teor.fis.uc.pt}, 
            Brigitte Hiller\footnote{Email address: 
            brigitte@teor.fis.uc.pt}}
\vspace{0.5cm}
\centerline{\small\it Centro de F\'{\i}sica Te\'{o}rica, Departamento de
         F\'{\i}sica da Universidade de Coimbra,}
\centerline{\small\it 3004-516 Coimbra, Portugal}
\vspace{1cm}

\centerline{\bf Abstract}
\vspace{0.5cm}
The six-quark instanton induced 't Hooft interaction, which breaks 
the unwanted $U_A(1)$ symmetry of QCD, is also a sourse of
semi-classical corrections to the low energy effective action. 
It is argued that there emerges a dimensionless expansion parameter 
that introduces a new mass scale, $\Lambda_\e^2\simeq 6\
\mbox{GeV}^2$, in the $0^-, 0^+$ channels. This scale plays a similar 
role as the large critical mass $M^2_\crit\gtrsim 4.2\div 6.6\ 
\mbox{GeV}^2$ discovered in the framework of QCD sum rules in the 
$0^-, 0^+$ gluonic channels. In particular, it allows to resolve the
$\eta'$ puzzle. To extract $\Lambda_\e$ we calculate 
the masses of the lightest pseudoscalar meson nonet by using the Nambu --
Jona-Lasinio (NJL) type $U_L(3)\times U_R(3)$ chiral symmetric 
Lagrangian together with the 't Hooft determinant. The mechanism 
which leads to the large value of $\Lambda_\e$ is scrutinized.

\newpage


\section*{\normalsize 1. Introduction}

Unfortunately, there is at present no quantitative framework within 
QCD to deal with its large distance dynamics. The physics of hadrons 
is approached through phenomenological parametrizations usually 
based on some simple ansatz with solid symmetry grounds. There are two 
important experimental facts to support this line of investigation.
First, it is known that the chiral symmetry of the massless QCD 
Lagrangian, which should be a good approximation for light quarks 
$(u,d,s)$, is not seen in the hadronic spectrum (the $SU(3)_V$ 
degenerate multiplets with opposite parity do not exist). It means 
that the hadronic vacuum is not symmetric under the chiral group. 
Second, it is seen from the mesonic spectrum that the $U_A(1)$ 
symmetry of the QCD Lagrangian is badly broken. The $SU(3)$ singlet 
pseudoscalar $\eta'$ is too heavy to be the ninth Goldstone boson. 
The $U_A(1)$ anomaly is responsable for the $\eta' -\pi ,K,\eta$ 
splitting \cite{Adler:1969}. It has been understood later  
\cite{Witten:1979a} that the $1/N_c$ expansion can be a 
relevant approximation to generate hadronic bound states and to find 
the singlet-octet splitting as a next to the leading $1/N_c$ order effect.   

A qualitatively correct picture of both spontaneous chiral symmetry 
breaking and $U_A(1)$ breaking at low energies, which is also compatible
with the conclusions coming from the large $N_c$ expansion, is given
by instantons \cite{Hooft:1976,Diakonov:1996}. The semi-classical theory 
based on the QCD instanton vacuum provides convincing evidences that 
$2N_f$-quark interactions ($N_f$ is the number of flavours) actually 
exist in QCD and that in the leading $1/N_c$ order they are described by 
the 't Hooft determinant \cite{Hooft:1978},
\begin{equation}
\label{Ldet}
  {\cal L}_{2N_f}=\kappa (\mbox{det}\ \bar{q}P_Lq
                         +\mbox{det}\ \bar{q}P_Rq)
\end{equation}
where the matrices $P_{L,R}=(1\mp\gamma_5)/2$ are projectors and the 
determinant is over flavour indices. We assume here that all 
interactions between quarks can be taken in the long wavelength limit 
where they are effectively local\footnote{To lowest order 
in $1/N_c$, hadronic physics can be described as a tree approximation 
to some local Lagrangian, with local hadron fields and local 
interaction vertices \cite{Witten:1979b}.}. At next to the leading 
$1/N_c$ order this vertex is modified by the tensor term which we have 
omitted in eq.(\ref{Ldet}). Even in this essentially simplified form the 
determinantal interaction contains all necessary features to describe 
the dynamical symmetry breaking of the hadronic vacuum and explicitly 
breaks the axial $U_A(1)$ symmetry. In the following we will asume
that quark fields have colour, $N_c=3$, and flavour, $N_f=3$, indices
which range over the set $i=1,2,3$. The coupling constant $\kappa$ is
a dimensional ($[\kappa ]=\mbox{GeV}^{-5}$) negative parameter 
with the large $N_c$ asymptotic $\kappa\sim 1/N_c^{N_f}$.  

On lines suggested by multicolour chromodynamics, however, it can 
be argued \cite{Witten:1979a} that the $U_A(1)$ anomaly is negligible in 
the large $N_c$ limit, the deviation of the singlet -- octet mixing 
angle from its ideal value is suppressed, so that mesons come
degenerate in mass nonets. Thus, the leading order mesonic Lagrangian 
must inherit the $U_L(3)\times U_R(3)$ chiral symmetry of massless
QCD. To specify the corresponding part of the effective quark 
Lagrangian we consider the four quark NJL type interections 
\cite{Nambu:1961}
\begin{equation}
\label{L4q}
  {\cal L}_\njl =\frac{G}{2}\left[(\bar{q}\lambda_aq)^2+
                 (\bar{q}i\gamma_5\lambda_aq)^2\right],
\end{equation}
where $\lambda_a,\ a=0,1,\ldots 8$ are the normalized ($\mbox{tr}
\lambda_a\lambda_b =2\delta_{ab})$ Gell-Mann matrices acting in
flavour space. The positive four quark coupling $G$,
$[G]=\mbox{GeV}^{-2}$, counts as $G\sim 1/N_c$ and therefore the 
Lagrangian (\ref{L4q}) dominates over ${\cal L}_6$ at large $N_c$. 
    
There is another approach to the $U_A(1)$ problem which is also based 
on the method of effective Lagrangians \cite{Vecchia:1980}. The way we 
follow in the present work reflects the quark structure of light 
pseudoscalar and scalar mesons and has a built-in mechanism for 
dynamical chiral symmetry breaking. A similar model has been
considered for two flavours in \cite{Bernard:1987} and for three 
flavours in \cite{Bernard:1988,Reinhardt:1988} and has been widely 
explored since that time \cite{Hatsuda:1994}. 

We have used the multicolour asymptotics to motivate our choice 
of many quark interactions (\ref{Ldet}) and (\ref{L4q}).
However, we are not going to follow explicitly the idea of $1/N_c$ 
expansion. As is well known from QCD sum rules, the channels with 
quantum numbers $J^P=0^+, 0^-$ are strongly coupled to the gluonic 
world \cite{Shifman:1981}. It has been argued there that for these 
channels the pictures emerging at $N_c\rightarrow\infty$ and $N_c=3$ 
seem to be qualitatively different from each other and the accuracy 
of the $1/N_c$ expansion becomes worse. From a pure phenomenological 
point of view it would suffice to mention here the large deviations from 
the Zweig rule in the pseudoscalar channel, or the well known $\eta'$ 
puzzle: the mass of this meson, being of order $1/N_c$, is 
unexpectedly too large. QCD sum rules relate these deviations from
the $1/N_c$ counting to the increasing of the mass scale $M_\crit$
characterizing the breaking of asymptotic freedom in the corresponding 
channel, in such a way that at large energies $E\gg M_\crit $, the
$1/N_c$ counting rehabilitates itself. The mass scale relevant for 
$\eta'$ physics is $M_\crit\gtrsim 2.0\div 2.6\ \mbox{GeV}$ 
\cite{Shifman:1981}.  

Since the critical mass is too large, one cannot rely much on the 
$1/N_c$ expansion. Here we use another approximation, namely the 
stationary phase method for the semi-classical path integral 
bosonization of the effective quark Lagrangian \cite{Reinhardt:1988} 
\begin{equation}
\label{totlag}
  {\cal L}=\bar{q}(i\gamma^\mu\partial_\mu -\hat{m})q
          +{\cal L}_\njl +{\cal L}_6.
\end{equation}
In this approximation the interactions ${\cal L}_\njl$ and 
${\cal L}_6$ are considered as contributions of the same order in
$\hbar$.

One can also try to improve the lowest order SPA result by taking
into account the Gaussian fluctuations of the six-quark 't Hooft 
determinant around the stationary phase trajectory 
\cite{Osipov:2002,Osipov:2004}. If ${\cal L}_6$ is not small, 
corrections can be much larger than one could predict starting 
from $1/N_c$, and be important for the mesonic ($0^+,0^-$) mass 
spectra. The dimensionless parameter ensuring the smallness of 
semi-classical corrections in the model is \cite{Osipov:2004}  
\begin{equation}  
\label{zeta}
   \zeta =\frac{\kappa^2 \Omega^{-1}}{32 G^3}\sim \frac{1}{N_c^3}
\end{equation}  
where $\Omega$ is the volume of a small Euclidean spacetime box with a 
side $2\pi /\Lambda_\e$. For certain, we are dealing here with a small
effect. However, one cannot neglect such quasi-classical contributions
until the mass scale $\Lambda_\e$ associated with it is established.
The strong $1/N_c$ suppression of (\ref{zeta}) makes room for a
large value of $\Lambda_E$ still leaving $\zeta$ small enough.
We argue here that the cut-off $\Lambda_\e$ is in a sense similar to the 
mass scale $M_\crit$ advocated in \cite{Shifman:1981}. Our assertion 
is based on a calculation of the mass spectrum of pseudoscalar $(0^-)$ 
mesons from which one can extract $\Lambda_\e$ and show that it is 
large $\Lambda_\e\gtrsim 2\ \mbox{GeV}$.   
   
It is worth noting that the model under consideration contains  
a second dimensionless parameter \cite{Osipov:2004}, 
\begin{equation}
\label{epsilon}
   \epsilon =\frac{|\kappa| \Delta}{4 G^2} \sim 
   \frac{1}{N_c} 
\end{equation}
with $\Delta=m-\hat m$, where $m$ stands for the constituent and $\hat
m$ for the current quark mass. The series expansion in $\epsilon$ 
closely corresponds to the $1/N_c$ expansion of the model. However, 
the fit to the meson mass spectrum shows that $\epsilon\simeq 0.7$, 
being in contradiction even with the $N_c=3$ estimate from 
(\ref{epsilon}). This relatively large (in comparison with $1$) value 
implies large $1/N_c$ corrections which convert the $1/N_c$ series 
into a badly convergent one. Sharing ideas of paper \cite{Shifman:1981}, 
we explain this behaviour of the series by the existence of a large 
critical mass in the channel. In this sense, the model perfectly 
reflects the known resolution of the $\eta'$ puzzle by generating 
quasi-classically a large mass scale parameter $\Lambda_E$ which 
makes $\eta'$ light on its natural scale 
$m_{\eta'}^2/\Lambda_\e^2\sim \zeta\sim 10^{-1}$. 


\section*{\normalsize 2. Characteristic scale of semi-classical 
                         corrections}  

Let us briefly recall some results of bosonization of the 
quark Lagrangian (\ref{totlag}). Important details can be found in 
\cite{Osipov:2004}. On the first stage one should linearize the many 
fermion vertices by introducing auxiliary bosonic fields. The pure quark
Lagrangian, ${\cal L}$, is transformed to a mixed meson-quark one
\begin{equation}
\label{Lmix}
   {\cal L}_\mix (q,\phi ,\sigma )
              ={\cal L}_q + {\cal L}_r + \Delta {\cal L}_r\ .  
\end{equation} 

The first term describes the tree level interactions of constituent 
quarks with pseudoscalar, $\phi_a (x)$, and scalar, $\sigma_a (x)$, 
$U(3)$ flavour nonets
\begin{equation}
   {\cal L}_q = \bar{q}(i\gamma^\mu\partial_\mu - m - 
                \sigma -i\gamma_5\phi )q.
\end{equation} 
   
The second term is the leading order stationary phase result 
\begin{eqnarray}
\label{Lr}
  {\cal L}_r\!\!\!\!\!\!\!\! 
          && = \frac{G}{12}\ \mbox{tr}\ (U_\st U^\dagger_\st )
             + \frac{1}{6}\ \mbox{tr}\ (WU_\st^\dagger + W^\dagger
               U_\st ) \nonumber \\
          && = h_a\sigma_a 
             + \frac{1}{2}h^{(1)}_{ab}\sigma_a\sigma_b
             + \frac{1}{2}h^{(2)}_{ab}\phi_a\phi_b + \ldots\ .
\end{eqnarray}
Here we used the stationary phase condition 
\begin{equation}
\label{spc}
   GU_a + W_a + \frac{3\kappa}{32}A_{abc}U^\dagger_bU^\dagger_c = 0
\end{equation} 
where the totally symmetric constants $A_{abc}$ are defined through 
the flavour determinant $\det W =A_{abc}W_aW_bW_c$. Our
notations are the following. The trace is taken over flavour indices, 
any flavour matrix written without open index is undestood as summed 
with the Gell-Mann $\lambda_a$ matrices ($a=0,1...8$), for instance, 
$W = W_a\lambda_a$. The mesonic fields are grouped in the covariant
combinations $W_a=\sigma_a +\Delta_a -i\phi_a$. The field 
$U_\st$ represents the exact solution of the stationary phase 
condition (\ref{spc}), which we seek in the form $U_a = s_a -ip_a$
expanding $s_a, p_a$ in increasing powers of bosonic fields 
$\phi_a, \sigma_a$. 
\begin{eqnarray}
\label{rsta}
   s_a^\st \!\!\!\!\!\!\!\!
      && = h_a+h_{ab}^{(1)}\sigma_b
           +h_{abc}^{(1)}\sigma_b\sigma_c
           +h_{abc}^{(2)}\phi_b\phi_c 
         + \ldots \\
   p_a^\st \!\!\!\!\!\!\!\!
      && = h_{ab}^{(2)}\phi_b
         + h_{abc}^{(3)}\phi_b\sigma_c
         + \ldots 
\end{eqnarray}
with coefficients $h^{(k)}_{ab...}$ explicitly depending on the quark 
masses and coupling constants $G,\kappa$. The coefficients are fixed by
the series of coupled equations following from (\ref{spc}) and
obtained by equating to zero the factors before independent 
combinations of mesonic fields. Due to recurrency of the considered 
equations all coefficients are determined once the first one, 
$h_a$, has been obtained \cite{Osipov:2004}.  

An alternative form to the exact solution of eq.(\ref{spc}) is
provided by the $1/N_c$ expansion which gives the stationary 
phase solution, $U_\st^a$, in form of the series 
\begin{equation}  
   U^\st_a = -\frac{1}{G}\left(
             W_a + \frac{3\kappa}{32G^2} A_{abc} W^\dagger_b
             W^\dagger_c + {\cal O}(1/N_c^2) \right),
\end{equation}
yielding for ${\cal L}_r$
\begin{equation}
\label{LrNc}
   {\cal L}_r = -\frac{1}{4G}\ \mbox{tr}\ (WW^\dagger )
                -\frac{\kappa}{(4G)^3}\left(
                \det W +\det W^\dagger \right)
                + {\cal O}(1/N_c).
\end{equation}
In fact, the large-$N_c$ limit forces a series expansion for the
coefficients of Lagrangian (\ref{Lr}) and this is how the dimensionless
parameter (\ref{epsilon}) reveals itself. To see this, let us assume 
for a moment that $SU_f(3)$ flavour symmetry is preserved. In this 
case the only coefficient among $h_a$ which is different from zero 
is $h_0$ and it is given by 
\begin{equation}
   h_0 = -\frac{8G}{\kappa} \sqrt{\frac{3}{2}}
         \left(1-\sqrt{1-\frac{\kappa\Delta}{4G^2}}\right).
\end{equation}    
If the ratio $\epsilon = \kappa\Delta /(4G^2)$ is small, one can 
expand the square root inside $h_0$ in powers of $\epsilon$. This
automatically leads to the same expansion for all other coefficients
in (\ref{Lr}) and, finally, one obtains Lagrangian (\ref{LrNc}) at
leading order in $\epsilon$. This limit is not affected by the 
semi-classical corrections $\Delta {\cal L}_r$, because these are
at most of order $\sim 1/N_c$. One can argue, however, that  
numerically $\epsilon\simeq 0.7$ (see our discussion in Sec.3) 
what is relatively large, to support the fast convergence of the 
series. 

The third term in (\ref{Lmix}) is the next to the leading order 
correction in the semi-classical expansion of the bosonized 
Lagrangian \cite{Osipov:2004}.  
\begin{equation}
\label{lagrfluc}
   \Delta {\cal L}_r= - \Omega^{-1} \sum_{n=1}^\infty 
                     \frac{(-1)^{n}}{2n}\ \mbox{tr}\ 
                     [F_{\alpha\beta} (\phi ,\sigma )]^n
\end{equation}
where 
\begin{equation}
\label{fab}
   F_{\alpha\beta}(\phi ,\sigma )=\frac{3\kappa}{16} A_{cbe}        
     \left(
     \begin{array}{cc}
     -h^{(1)}_{ac}{\bar s}_e^\st & h^{(1)}_{ac}p_e^\st \\
     h^{(2)}_{ac}p_e^\st         & h^{(2)}_{ac}{\bar s}_e^\st
     \end{array}
     \right)_{\alpha\beta},  
\end{equation}
with ${\bar s}_a^\st = s_a^\st - h_a$. The factor $\Omega^{-1}$ may
be written as an ultraviolet divergent integral regularized by 
introducing a cut-off $\Lambda_\e$\footnote{One should not confuse this 
parameter with the standard ultraviolet NJL cut-off $\Lambda\simeq 1\ 
\mbox{GeV}$ for the quark loops, which represents the mass scale of 
spontaneous chiral symmetry breaking and determines the value of the 
quark condensate to leading order.} 
\begin{equation}
   \Omega^{-1} = \delta_\e^4(0)\sim 
                 \int_{-\Lambda_\e /2}^{\Lambda_\e /2}
                 \frac{\ud^4k_\e}{(2\pi )^4}=
                 \frac{\Lambda^4_\e}{(2\pi )^4}\ .
\end{equation}  

One can speculate about the value of the cut-off. By definition
$\Lambda_\e$ belongs to the ``quark territory'' $\Lambda_\e\gtrsim 1\ 
\mbox{GeV}$. On the other side, only effects that go beyond 
standard perturbation theory are included into the Lagrangian 
${\cal L}_6$. Therefore, the characteristic volume of quantum fluctuations, 
$\Omega$, can not deviate much from the size determined by 
non-perturbative fluctuations corresponding to classical solutions of 
the non-linear Yang-Mills equations, i.e. instantons. The 
relevant mass scale is generated by the gluon vacuum condensate. 
Thus, we have as a crude estimate 
\begin{equation}  
   \Omega\simeq \big <0|\frac{\alpha_s}{\pi}
               G^a_{\mu\nu}G^a_{\mu\nu}|0\big >^{-1} 
               \simeq (330\ \mbox{MeV})^{-4}\ , \qquad 
   \Lambda_\e\simeq 2.1\ \mbox{GeV}.  
\end{equation}
 
The crucial question is, however, whether this large value is in 
agreement with the general idea of a semi-classical expansion: 
corrections must be small in comparison with the leading order 
result. At first sight, it seems that we have just the opposite 
case. The factor $\Omega^{-1}\sim \Lambda_\e^4$ and one can naively 
expect that large values of $\Lambda_\e$ will severely break the 
convergence of the quasi-classical series. 

This is not entirely true. To clarify the point and to learn one 
important feature of Lagrangian (\ref{lagrfluc}), let us use again the 
$1/N_c$ expansion. The first terms of (\ref{lagrfluc}) in this 
framework are 
\begin{equation}
\label{scc}
   \Delta {\cal L}_r = -\frac{\kappa^2\Omega^{-1}}{8(2G)^4}
          \left[\mbox{tr}\ (WW^\dagger )
                +\frac{3\kappa}{(4G)^2}\left(
                \det W +\det W^\dagger \right)
                + {\cal O}(1/N_c^2)\right].
\end{equation}
One can make here a few interesting observations. First, the
correction to the result (\ref{LrNc}) starts from a term of   
$1/N_c^2$ order. Due to fine cancellations in (\ref{lagrfluc})
it is two orders less than one would naively expect. Second, every 
term in eq.(\ref{scc}) is suppressed by the factor $\zeta$ (for its 
definition see eq.(\ref{zeta})) as compared with corresponding terms 
in (\ref{LrNc}). This dimensionless parameter measures the size  
of the semi-classical contribution. Its value must be small, but not 
necessarily so much suppressed as it follows from the above estimations. 
Actually, nothing forbids us to suppose that such a suppression is 
partly compensated due to the increase of the mass scale $\Lambda_\e$, 
in such a way that $\zeta$ still remains small, $\zeta\sim 10^{-1}$. 


\section*{\normalsize 3. A model estimation of $\Lambda_\e$ }  
  
Let us see how our expectations look numerically. To check our guess 
one should turn to the calculation of the pseudoscalar mass spectrum and 
extract $\Lambda_\e$ by confronting model results with experimental
data. To make our consideration not too overfilled with details of
numerical calculations we give here only final results. We postpone 
the details until a future publication \cite{Osipov:2005b}. 

In the following we shall consider the case of $SU(2)_I\times U(1)_Y$ 
symmetry, i.e. we take ${\hat m}_u={\hat m}_d\ne {\hat m}_s$. Accordingly, 
there are alltogether six parameters, $\hat{m}_u$, $\hat{m}_s$, $G$, 
$\kappa$, $\Lambda$ and $\Lambda_\e$, shown in Table 1. The last two 
collumns present the dimensionless parameters of the model, $\zeta$ and 
$\epsilon$, related to the different types of power expansions. In Table 
2 are the results of our calculations of the pseudoscalar spectrum, 
together with the weak decay constants $f_\pi$, $f_K$ and mixing angle 
$\theta_\p$ in the singlet -- octet basis ($\phi_0,\phi_8$). Inputs are 
indicated by (*). The Latin letter labels on the left hand side identify 
the sets in the tables.
\vspace{0.5cm}

\noindent\textsc{table 1.} {\small
Main parameters of the model given in the following units: 
$[m]=\mbox{MeV}$, $[G]=\mbox{GeV}^{-2}$, $[\kappa ]=\mbox{GeV}^{-5}$, 
$[\Lambda ]=\mbox{GeV}$. Sets $(a,b)$ correspond to the leading order
SPA. Sets $(c,d)$ include semi-classical corrections. Constituent 
quark masses $m_i$ and corresponding leading order results 
${\stackrel{\circ} m}_i$ are also given.} 
\begin{center}
\begin{tabular}{l ccc ccc c c c c c c c c}
\hline
\hline
      & $\hat{m}_u$ 
      & ${\stackrel{\circ} m}_u $ 
      & $m_u $ 
      & $\hat{m}_s$
      & ${\stackrel{\circ} m}_s $
      & $m_s $
      & $G$  
      & $-\kappa $ 
      & $\Lambda $
      & $\Lambda_\e $
      & $\zeta$ 
      & $\epsilon$   \\ 
\hline
a &5.3 &315 &-   &170 &513 &-   &8.89 &687  &0.92 &-   &-     &0.67 \\ 
b &6.1 &380 &-   &185 &576 &-   &12.6 &1116 &0.83 &-   &-     &0.66 \\ 
c &4.1 &446 &381 &123 &598 &559 &7.14 &201  &1.1  &2.4 &0.076 &0.44 \\ 
d &4.5 &471 &412 &132 &620 &585 &8.81 &298  &1.03 &2.3 &0.076 &0.45 \\
\hline
\hline
\end{tabular}
\end{center}   
\vspace{0.5cm}

The two first sets $(a,b)$ are obtained in the framework of leading 
order SPA, along lines suggested in our recent work 
\cite{Osipov:2005}. The quantum fluctuations given by the Lagrangian 
$\Delta {\cal L}_r$ are not taken into account. One can see that  
reasonable fits to the pseudoscalar spectrum at leading order 
correspond to a large value of  $\epsilon\simeq 0.7$ and clearly 
show the slow convergence of the $1/N_c$ series.
\vspace{0.5cm}

\noindent\textsc{table 2.} {\small 
The light pseudoscalar nonet characteristics (in units of MeV, 
except for the angle $\theta_\p$, which is given in degrees)
are presented in a full correspondence with the parameter sets
of Table 1.}
\begin{center}
\begin{tabular}{l c c c c c c c c c}
\hline
\hline
   &$-\big <\bar{u} u\big >^{1/3}$
   &$-\big <\bar{s} s\big >^{1/3}$
   &$m_\pi$
   &$m_K  $
   &$f_\pi$
   &$f_K  $
   &$m_\eta $
   &$m_{\eta'}$
   &$\theta_\p\ $ \\ 
\hline
a &244  &204 &138* &494* &92* &121*   &487  &958* &-12.0 \\
b &233  &182 &138* &499* &92* &115.8* &477  &958* &-15.0 \\
c &287* &263 &138* &499* &92* &115.8  &503* &958* &-12.5 \\ 
d &274* &244 &138* &494* &92* &113*   &494  &958* &-13.3 \\ 
\hline
\hline
\end{tabular}
\end{center}
\vspace{0.5cm}

In the sets $(c,d)$ we include the semi-classical correction to the
leading order result. In these cases a new mass scale parameter
$\Lambda_\e$ enters the fitting process. Consequently, we have 
included the value of the light quark condensate as an additional input. 
One can see that quantum fluctuations lead to a small effect, slightly 
improving the fit. Nevertheless, our expectations seem to be realized. 
The cut-off $\Lambda_\e$ has a large value 
$\Lambda_\e\simeq 2.4\ \mbox{GeV}$, with $\zeta\simeq 0.08$ 
being small enough.    

Let us try to understand why $\Lambda_\e \simeq 2.4\ \mbox{GeV}$
and not much larger. The reason for this is very simple and is contained
in the value for $\epsilon \simeq 0.44$. This value indicates that 
cancellations in $\Delta {\cal L}_r$ are not so strong as one would
expect by using the large-$N_c$ arguments to obtain (\ref{scc}).  
Being $35\%$ less, in comparison with sets $(a,b)$, the model parameter 
$\epsilon$ is still too large to vouch for suppression in Lagrangian
$\Delta {\cal L}_r$ in full measure. 

On the other hand, the calculations
would become unreliable if the value of $\Lambda_\e$ would lead to such
large corrections that they could compete with the leading term. We
have found that at $\Lambda_\e\simeq 2.4\ \mbox{GeV}$ the correction 
amounts at most to $\sim 20\%$ of the leading term (in the case of
weak decay constants $f_\pi$ and $f_K$). This is a strong signal that 
we are already quite close to exhaust the reserves of the
quasi-classical expansion and that a further increase of $\Lambda_\e$ 
would start to destroy the fast convergence of the series, giving, 
in addition, a worse fit.



\section*{\normalsize 4. Discussion of the result}  
  
We have pointed out a new interesting feature related with the 
't Hooft determinant resolution of the $U_A(1)$ problem, which provides 
one more argument in favour of this six-quark interaction: it also
unravels the obviously nontrivial $\eta'$ puzzle. Indeed,
the phenomenological consequences of the 't Hooft interaction are
well-known. For instance, it leads to spontaneous chiral symmetry 
breaking with the characteristic mass scale $\Lambda\sim 1\ \mbox{GeV}$ 
extracted from the quark loops. It also explains the deviations from 
Zweig's rule. In accordance with standard $N_c$ counting rules, 
the six-quark interaction yields a flavour singlet $\eta'$ meson mass 
of order $m_{\eta'}^2\sim 1/N_c$, as it is commonly expected. 
However, it has not been clear why this $1/N_c$ suppressed  
mass is not much smaller than its actual value of almost one GeV. The
mass scale for chiral symmetry breaking, $\Lambda$, is too low to 
explain this phenomenological fact. 

To find the answer we have suggested to bosonize the quark determinantal 
interaction and to take into account the next to the lowest order term 
in the semi-classical expansion of the bosonized 't Hooft Lagrangian. 
This formal expansion in powers of the dimensional parameter $\hbar$
actually contains a small dimensionless parameter $\zeta$, which hides
the large characteristic scale $\Lambda_\e\simeq 2.4\ \mbox{GeV}$.
   
Our solution is obtained in the framework of a simple model, although
it is strongly based on the instanton picture of the QCD vacuum. 
Accumulating the most essential features of the instanton physics,
it is not an accident at all, that it resolves the $\eta'$ puzzle in a 
very similar way to the known solution in the framework of QCD sum 
rules. There are, however, important differences: in the latter approach 
the $\eta'$ gets its mass through mixing with glue states and these 
are expected to be much heavier, giving a new large mass scale 
parameter for the pseudoscalar channel. 

The 't Hooft determinantal interaction is a remnant of gluodynamics at
large distances, i.e. at scales where quarks interact with each other   
through their zero modes in the instanton background. By means of the 
semi-classical expansion of the bosonized 't Hooft Lagrangian we are 
able to ``touch'' the border of the non-perturbative region from its  
low energy side, as opposed to the QCD sum rules method. Amusingly, 
the numerical result is in a perfect agreement with our expectations 
and with the value obtained on the basis of QCD sum rules.     

\section*{Acknowledgements}
This work has been supported by grants provided by Funda\c c\~ao para
a Ci\^encia e a Tecnologia, POCTI/35304/FIS/2000 and Centro de F\ii sica
Te\'orica unit 535/98.


\end{document}